\documentclass[aps,pra,reprint]{revtex4-1}

\usepackage{graphicx}
\usepackage{amsmath}

\begin{document}

\title{Raman cooling imaging: Detecting single atoms near their ground state of motion}

\author{B. J. Lester}
\email[E-mail: ]{blester@jila.colorado.edu}
\author{A. M. Kaufman}
\author{C. A. Regal}
\email[E-mail: ]{regal@colorado.edu}
\affiliation{JILA, National Institute of Standards and Technology and University of Colorado, and
Department of Physics, University of Colorado, Boulder, Colorado 80309, USA}

\begin{abstract} 
We demonstrate imaging of neutral atoms via the light scattered during continuous Raman sideband cooling.  We detect single atoms trapped in optical tweezers while maintaining a significant motional ground-state fraction. The techniques presented provide a framework for single-atom resolved imaging of a broad class of atomic species.
\end{abstract}

\date{\today}

\maketitle

The ability to resolve single atoms has been integral to many advances in quantum information processing, quantum optics, and quantum many-body physics~\cite{Wineland1998,Olmschenk2009,Hofmann2012,Schellekens2005,Beugnon2006,Gericke2008,Bakr2009,Weitenberg2011}. Imaging single atoms, such as trapped atomic ions or neutral atoms in optical lattices, is typically achieved by scattering near-resonant photons, a fraction of which are collected to generate a signal on a photodetector. The use of high numerical-aperture (NA) optics and single-photon counting detectors can reduce the required number of scattered photons~\cite{Fuhrmanek2011,Gibbons2011}, but it is often advantageous to scatter many photons to achieve a high signal-to-noise ratio (SNR). Such detection is especially useful to minimize the readout error for site-resolved, {\it in situ} imaging of multiple traps.

For neutral atoms, which do not benefit from the deep traps achieved in ion experiments, imaging requires maintaining sufficiently low temperatures to preserve the atomic configuration. A common technique for position-resolved detection of optical wavelength-spaced neutral atoms is  sub-Doppler cooling~\cite{Schlosser2001,Bakr2009,Sherson2010,Nelson2007}. However, this technique requires a favorable atomic level structure to work efficiently. Raman sideband cooling is an alternative imaging method~\cite{Nelson2007,Setiawan2012thesis} that is possible for any atomic species that can be trapped in a tightly-confining potential and has an electronic dark state. Alternatives to traditional sub-Doppler cooling are especially important for atoms where the hyperfine structure is not well resolved, such as lithium or potassium~\cite{Landini2011,Fernandes2012}. The extension of site-resolved imaging of quantum gasses to lighter atoms will enable quantum gas microscope experiments where dynamics occur on faster time-scales \cite{Setiawan2012thesis,Zurn2012,Yefsah2013}. Further, Raman sideband cooling is an established method for cooling an atom to the three-dimensional (3D) motional ground state~\cite{Monroe1995,Kerman2000,Han2000,Li2012,Kaufman2012} and hence provides the control to prevent thermal hopping in traps that support only a small number of localized bound states~\cite{Nelson2007}.  However, to date there has been no demonstration of the feasibility of the experimental detection of single atoms via Raman cooling.

\begin{figure}[hb!]
	\begin{center}
		\includegraphics[width=\columnwidth]{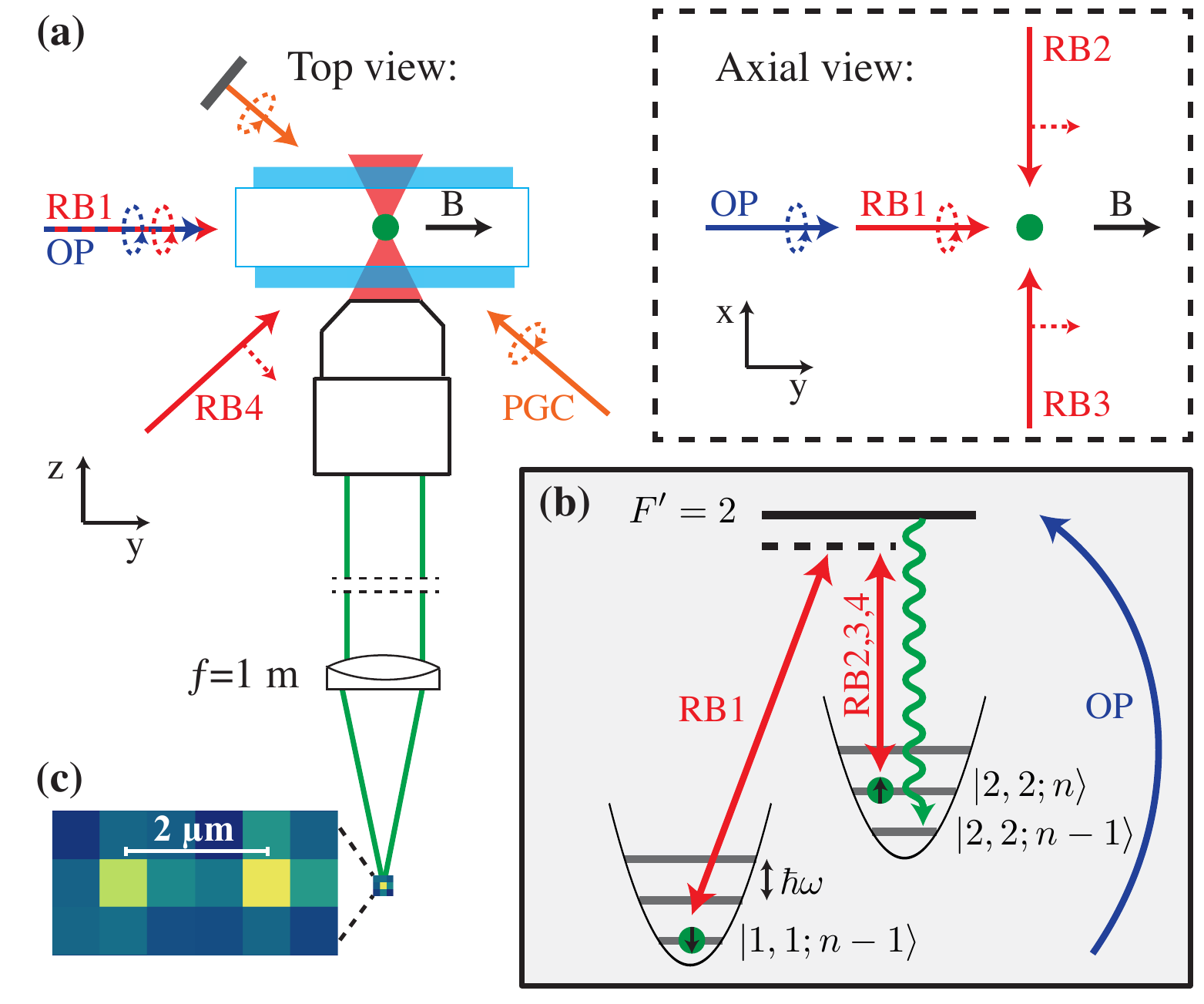}
		\caption{(Color online) 
		(a) Schematic of the experimental setup from above (Top view) and in the plane perpendicular to $\hat{z}$ (Axial view). The Raman beams (RB1,2,3,4; red), optical pumping beams (OP; blue), polarization-gradient cooling beams (PGC; orange), and the quantization axis (B; black) are indicated in relation to the optical tweezer potential. The dashed lines indicate the beam polarizations. The light scattered by the atom is collected by the same objective lens used to generate the optical tweezers and is imaged using an $f=1$ m achromatic doublet. 
		(b) Raman cooling cycle in 1D. Starting in the state $|F,m_F; n_\text{vib}\rangle = |2,2; n\rangle$, a coherent Raman sideband transition to the $|1,1;n-1\rangle$ state is performed using a pair of Raman beams (RB1 + RB2,3,4). Subsequently, a spontaneous Raman transition optically pumps the atom back to the $|2,2;n-1\rangle$ state. By repeating this process, atom population will accumulate in the $|2,2;0\rangle$ state, which is ideally a dark state.
		(c) Example of an image generated via continuous Raman cooling. The scale bar (white) indicates separation between tweezer traps for imaging.
		} 
		\label{fig:schematic}
	\end{center}
\end{figure} 

Here, we report the imaging of single $^{87}\text{Rb}$ atoms via the light scattered during a period of continuous Raman sideband cooling in an optical tweezer trap.  The efficiency of the cooling maintains a significant ground-state fraction while scattering photons. The operational principle of sideband cooling relies on the existence of a dark state, composed of a spin degree of freedom and the motional ground state, that is decoupled from the cooling light. Thus, an extended period of imaging requires some method by which the dark state is tunably compromised to allow continuous photon scattering. We therefore parametrically modulate the trapping potential to couple the atom out of the dark state during imaging. By varying the parameters of the drive, we can increase the detection fidelity and speed of the imaging procedure when a smaller ground-state fraction can be tolerated.

Figure~\ref{fig:schematic}(a) shows the optical tweezer apparatus in which we implement imaging via Raman cooling [Fig.~\ref{fig:schematic}(b)]~\cite{Kaufman2012,Diedrich1989,Hamann1998,Perrin1998,Forster2009}. The efficiency of the cooling cycle is determined by our ability to optically pump back to the initial spin state while preserving the reduced motional state, which is related to the Lamb-Dicke parameter $\eta^{OP} = x_0 k$, where $x_0=\sqrt{\hbar / 2 m \omega}$ is the oscillator length for a particle of mass $m$ with trap frequency $\omega$, and $k$ is the wavenumber of the optical pumping (OP) light. For our optical tweezer traps in this work, $U/k_B=1.1\text{ mK}$ and the radial (axial) trap frequency is $\omega_\text{r}=2 \pi\times140 \text{ kHz}$ ($\omega_\text{a}=2 \pi\times30 \text{ kHz}$), which gives $\eta^{OP}_\text{r}=0.16$ ($\eta^{OP}_\text{a}=0.35$)~\cite{Kaufman2012}. This work uses continuous Raman cooling, where the OP beams remain on during the entire procedure~\cite{Kaufman2014}. 

Each experimental cycle begins with a single atom that has been stochastically loaded by overlapping a magneto-optical trap (MOT) with an optical tweezer trap~\cite{Schlosser2002,Sompet2013}. After 10 ms of polarization-gradient cooling (PGC), the atom is cooled to near the bottom of the trap, with a final temperature of $\sim15\;\mu\text{K}$~\cite{Dalibard1989,Tuchendler2008,Kaufman2012}. The trap occupancy is then measured with a standard PGC fluorescence image~\cite{Schlosser2001,Kaufman2012}, which works very well for $^{87}\text{Rb}$. We use this as a pre-selection image to indicate the presence of an atom for the studies we present.  After the pre-selection image, two more images are taken in each experimental cycle. The second image is the exposure during continuous Raman cooling; this image provides the main signal for this work. The final image is another PGC fluorescence image (the post-selection image) that is used in combination with the pre-selection image to distinguish experiments where the atom was lost. 

\begin{figure}[tb!]
	\begin{center}
		\includegraphics[width=\columnwidth]{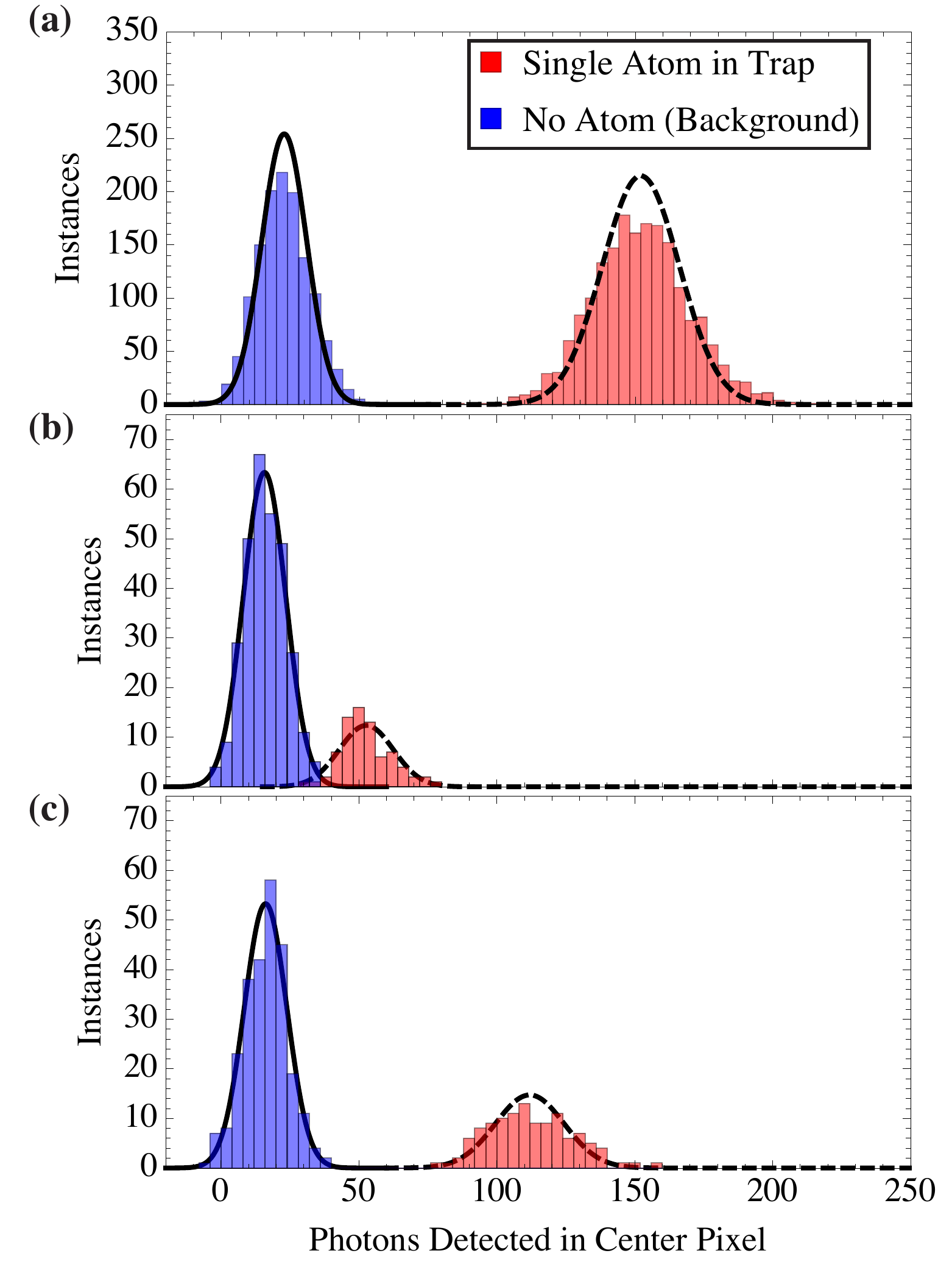}
		\caption{(Color online) 
		Photon detection histograms. The red (blue) bars on the right (left) indicate images taken with (without) an atom in the trap, as determined by the pre-selection image. The dashed (solid) black gaussian curve approximates the expected distribution for the atomic signal (background).
		(a) Fluorescence signal from 25 ms of PGC imaging. 
		(b) Signal collected during 1 s of continuous Raman cooling. The presence of an atom is triggered using the pre- and post-selection PGC images.
		(c) same as in (b), but now applying a sinusoidal parametric drive with $\Delta U/U\sim 0.055$, resonant with the radial trap frequency, to the optical tweezer potential. 
		}
		\label{fig:histograms}
	\end{center}
\end{figure} 

For all of our data analysis, the recorded signal is the number of photons detected in a single pixel on which the atomic signal is centered [Fig.~\ref{fig:schematic}(c)]. This represents $\sim1/3$ of the total detected signal. While the best SNR is achieved when all of the signal is binned onto a single pixel, we operate the camera with a smaller pixel size to fully separate the signal from atoms in different traps [Fig.~\ref{fig:schematic}(c)]. However, for absolute comparison of the measured scattering rates (discussed below), we integrate the signal on the surrounding 3$\times$3 pixels to ensure that the majority of detected photons are counted. 

Figure~\ref{fig:histograms}(a) shows an example of the signal from 25 ms of PGC fluorescence imaging, which provides a useful benchmark for imaging single neutral atoms~\cite{Schlosser2001,Nelson2007}.  For this comparison, we use a single pair of PGC beams in the $\sigma ^+\text{-}\sigma^-$ configuration [shown in Fig.~\ref{fig:schematic}(a)] with 0.61 mW of power in a beam with a gaussian waist of 1.66 mm~\footnote{Note that the pre- and post-selection images used a six-beam PGC configuration, which we find achieves a comparable SNR.}. This configuration minimizes interferometric intensity gradients near the atom, which reduces fluctuations in the atomic scattering rate~\cite{Dalibard1989}. We red detune the PGC light 38 MHz from the trap-shifted cycling transition, which corresponds to $13$ MHz from the free-space transition~\cite{LeKein2013,Safronova2011,Arora2012,Shih2013}. Taking into account the $6\%$ collection efficiency of our optical system, the measured scattering rate during PGC imaging is $\left(2.6\pm0.4\right)\times10^5\text{ photons s}^{-1}$. This is reasonable in comparison to the estimate of $4.5\times10^5\text{ photons s}^{-1}$\cite{Steck2010}, which assumes perfect alignment of the peak intensity of the PGC beams to the atom, balanced circularly-polarized beams, and the ideal steady-state spin distribution for an atom at rest in one-dimensional (1D) PGC light~\cite{Dalibard1989}.

During Raman cooling imaging, 3D cooling must be performed even if our controlled parametric drive affects only one motional axis. This requirement arises from the optical pumping process, which can heat any motional axis via spontaneous emission. We use three beams to simultaneously perform continuous Raman cooling along both the radial (RB1+RB2) and axial (RB1+RB4) dimensions of the trap. For the radial cooling, we take advantage of the near degeneracy of the two radial trap frequencies of our trap to cool with only one beam pair and still achieve the necessary (albeit slower) cooling along both dimensions. 

Figure~\ref{fig:histograms}(b) shows a histogram of the number of photons detected during 1 s of continuous Raman sideband cooling. The red (blue) data represent experiments where an atom was present (absent) in both the pre- and post-selection images, from which we infer that there was an atom present (absent) during the entire period of continuous Raman cooling. Selecting experiments where the atom is not lost allows us to directly characterize the imaging procedure, independent of imperfections (such as the vacuum lifetime) that are specific to our setup. We note that the signal peak is smaller than the background peak because the trap loading rate was smaller than experiments in Fig.~\ref{fig:histograms}(a) (roughly 25\% instead of 60\%) due to a lower MOT density during loading. This does not affect the results other than to reduce the available statistics for the signal data points compared to background. 

In principle, we should expect no signal when no parametric drive is applied to the trap because the atom should remain in the dark state and thus scatter no photons. The main contributions to the signal in Fig.~\ref{fig:histograms}(b) are from off-resonant carrier [$\Delta n$=0 in Fig.~\ref{fig:schematic}(b)] Raman transitions, which initiate an optical pumping cycle, and, to a lesser extent, residual coupling of the dark state to the optical pumping beams.  Off-resonant carrier transitions are relevant in these experiments because the ratio of the carrier Rabi rate (the transition linewidth) to the trap frequency (the detuning from the transition) is non-negligible, leading to a finite population transfer. The measured scattering rate for the data in Fig.~\ref{fig:histograms}(b) is $(1.9\pm0.3)\times 10^3$ photons s$^{-1}$. We estimate the maximum scattering rate from off-resonant carrier transitions to be $2 \times 10^3 \text{ photons s}^{-1}$ by summing the scattering rates from two 1D calculations of the axial and radial rates. 

The fidelity of the imaging is determined by a combination of the signal amplitude and the ability to separate the signal and background distributions.  The expected standard deviation in the number of detected photons is a combination of shot noise (Poisson statistics) and  the readout noise of our detector ($\sigma_{RO}=6.6 \;\frac{\text{photons}}{\sqrt{\text{pixel}}}$). The expected distributions for the atomic signal (background) are approximated by the dashed (solid) gaussian curves on the histograms in Fig.~\ref{fig:histograms}.  Based on the overlap integral of two normalized gaussian distributions with means and standard deviations from the data in Fig.~\ref{fig:histograms}(b), we estimate a minimum detection error rate of $1.4\times10^{-2}$ for 1 s of integration time~\cite{Fuhrmanek2011}~\footnote{The minimum detection error rate $\delta$ is half the overlap integral of the two normalized gaussian distributions, which requires an optimal choice of threshold to achieve. The fidelity of the measurement is then defined to be $\mathcal{F}=1-\delta$ (i.e., $\delta=\epsilon_B=\epsilon_D$ in Ref. \cite{Fuhrmanek2011}).}. Note that this is a measure of the achievable SNR, but the effects of atom loss are removed based on the pre- and post-selection images.  

To improve the signal detected during continuous Raman cooling, we apply a parametric drive by modulating the total optical power generating the optical tweezer trap at twice the radial trap frequency ($2 \omega_\text{r}=2\pi\times280$ kHz). This drives the atom out of the motional ground state, allowing the cooling cycle to restart, and thus more OP photons are scattered during the exposure. Figure~\ref{fig:histograms}(c) clearly shows an increase in the atomic signal during Raman cooling imaging with a parametric drive applied to the trapping potential. The increased signal corresponds to a reduction of the minimum detection error rate to $<10^{-4}$ for the same 1 s integration time.  For comparison, we estimate that 20 ms of our PGC imaging, which has a much higher scattering rate, can achieve the same minimum detection error rate.

\begin{figure}[tb!]
	\begin{center}
		\includegraphics[width=\columnwidth]{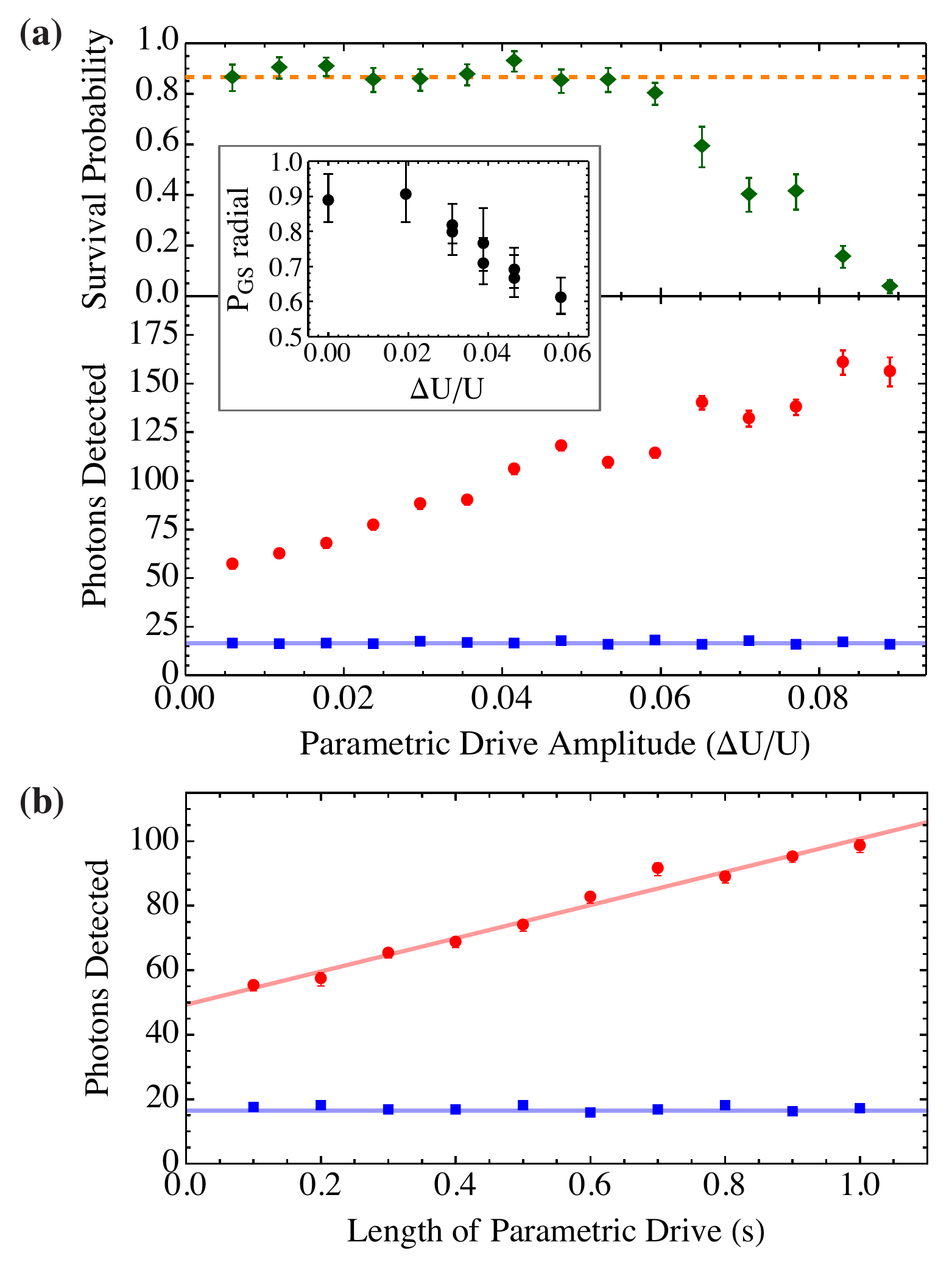}
		\caption{(Color online) 
		(a) Varying the amplitude of the parametric drive applied to the optical tweezer potential during 1 s of Raman cooling imaging.
		 The green diamonds are the atom survival probability, as determined using the pre- and post-selection images, and the orange dashed line is the measured vacuum lifetime limit. The red circles (blue squares) are the signal collected with (without) an atom in the trap during the entire imaging period. The blue line indicates the average background value of 16.5 detected photons.  
		 The inset displays the equilibrium radial ground-state fraction ($P_{GS}$ radial) measured via Raman sideband spectroscopy immediately following a period of Raman cooling imaging \cite{Kaufman2012}.
		(b) Signal as a function of the parametric drive length, for fixed total imaging length (1 s) and drive amplitude ($\Delta U/U\sim0.04$). The red line is a linear fit to the data.  All error bars indicate the standard error of the measurement.
		}
		\label{fig:paramData}
	\end{center}
\end{figure} 

The Raman cooling imaging procedure can be simply understood as the competition between a variable parametric excitation rate and the (constant) continuous Raman cooling. The competition of these rates rapidly yields an equilibrium ground-state fraction, and hence non-dark fraction, which largely determines the measured scattering rate. This simple description is effective because the parametric drive does not directly interfere with the cooling and the Raman cooling cycle occurs on sub-millisecond timescales compared to the 1 s image. 

The average number of photons detected during Raman cooling imaging is observed to increase as a function of the applied parametric drive amplitude [red circles in Fig.~\ref{fig:paramData}(a)]. Hence, increasing parametric drive amplitude reduces the equilibrium ground-state fraction, allowing us to control the atomic fluorescence rate. The atom signal can be compared to the average number of background photons detected (blue squares) measured at each drive amplitude.  The amplitude of the parametric drive is defined as the ratio of the peak-to-peak variation of the trap depth $\Delta U$ to the trap depth in the absence of a drive $U$.  

Up to intermediate drive amplitudes, we characterize the equilibrium radial ground-state fraction during imaging by performing radial sideband thermometry immediately after a period of Raman cooling imaging. We verify that the equilibrium ground-state fraction varies with the applied parametric drive amplitude [inset of Fig.~\ref{fig:paramData}(a)].  For the lowest drive amplitudes ($\Delta U/U\lesssim0.02$), thermometry is limited by the resolution of our sideband spectroscopy~\cite{Kaufman2012}, suggesting a radial ground-state fraction $P_{GS}>80\%$. We further find that a significant ground-state fraction $P_{GS}\ge55\%$ is maintained for drive amplitudes up to $\Delta U/U\sim0.06$ (near the maximum drive without loss). The scattering rate observed at this drive amplitude is $\left(4.9\pm 0.7\right) \times 10^3$ photons s$^{-1}$. We estimate an expected scattering rate of $\le 7 \times 10^3 \text{ photons s}^{-1}$ by weighting the scattering rates from each of the motional excited states undergoing Raman cooling, assuming a thermal distribution corresponding to the measured motional occupation at $\Delta U/U\sim0.06$~\cite{Kaufman2012}.

Further, the survival probability of the atom is measured as a function of the parametric drive amplitude [green diamonds in Fig.~\ref{fig:paramData}(a)] by recording the probability for an atom to appear in the post-selection image, given that it was present in the pre-selection image.  The orange dashed line represents the expected survival probability due to the vacuum lifetime ($\tau=7.1$ s) over the course of the 1.023 s between the pre- and post-selection images.  The atom loss is initially consistent with background collisions, but at larger drive amplitudes the parametric excitation rate exceeds the Raman-cooling rate and the atom is quickly heated out of the trap.  

In Fig.~\ref{fig:paramData}(b), we perform 1 s of continuous Raman cooling and apply a parametric drive for a fraction of the image. The constant drive amplitude of $\Delta U/U\sim0.04$ realizes an equilibrium radial ground-state fraction of $\ge 70\%$ [inset of Fig.~\ref{fig:paramData}(a)]. 
The measured data (red circles) are well fit by a linear increase (red line), which is further evidence that the radial ground-state fraction rapidly reaches equilibrium. For the same drive amplitude, we estimate that a minimum detection error of $<10^{-2}$ can be reached after about 0.5 s of imaging. This is comparable to the site-resolved imaging used in recent optical lattice experiments \cite{Bakr2009,Sherson2010}.

For atomic species where sub-Doppler cooling is not sufficient for imaging, such as desired experiments with $^6$Li in an optical lattice, continuous Raman cooling is a promising technique for {\it in situ} imaging \cite{Setiawan2012thesis}.  For moderate lattice depths, such that the trap frequency for lithium is $\omega_\text{Li} \sim 2 \pi \times1$ MHz, the Lamb-Dicke parameter becomes $\eta_{\text{Li}}^{OP}=0.27$. Efficient sideband cooling has been demonstrated for comparable Lamb-Dicke parameters \cite{Forster2009,Li2012,Kaufman2012}. Thus, the confinement of lithium on a lattice should not be a fundamental limitation to performing continuous Raman sideband cooling for imaging. The remaining concern is whether the number of photons scattered during optical pumping is small enough to preserve the reduced motional state in each cooling cycle; however for imaging (not ground-state cooling), even this might be overcome by addressing the second (or higher) order sidebands. 

In conclusion, we have demonstrated the position-resolved detection of single neutral atoms using Raman sideband cooling.  In our studies we use $^{87}\text{Rb}$, for which PGC fluorescence imaging is an order of magnitude faster than Raman cooling imaging; hence PGC will remain an imaging method of choice for many applications with heavy alkali atoms. However, Raman cooling imaging adds an important technique for preserving ultracold temperatures and hence atomic configurations in shallow potentials.  Further, we believe Raman cooling imaging will be imperative for the site-resolved imaging of lighter atoms that do not benefit from the well-resolved level structure of atoms such as $^{87}\text{Rb}$.  For one- or two-dimensional systems (as well as sparse 3D systems), where rescattering effects are suppressed \cite{Wolf2000}, these results can be applied to larger clouds of atoms, as long as the trapping potential provides sufficient confinement. \\

{\em Note} -- During preparation of the manuscript, we learned of a related work in Ref.~\cite{Patil2014}.\\

We thank T. P. Purdy, C. M. Reynolds, and K. R. A. Hazzard for useful  discussions. This work was supported by the David and Lucile Packard Foundation and the National Science Foundation under grant number 1125844.  CAR acknowledges support from the Clare Boothe Luce Foundation, BJL from the NSF-GRFP, and AMK from NDSEG.


%


\begin{thebibliography}{41}%
\makeatletter
\providecommand \@ifxundefined [1]{%
 \@ifx{#1\undefined}
}%
\providecommand \@ifnum [1]{%
 \ifnum #1\expandafter \@firstoftwo
 \else \expandafter \@secondoftwo
 \fi
}%
\providecommand \@ifx [1]{%
 \ifx #1\expandafter \@firstoftwo
 \else \expandafter \@secondoftwo
 \fi
}%
\providecommand \natexlab [1]{#1}%
\providecommand \enquote  [1]{``#1''}%
\providecommand \bibnamefont  [1]{#1}%
\providecommand \bibfnamefont [1]{#1}%
\providecommand \citenamefont [1]{#1}%
\providecommand \href@noop [0]{\@secondoftwo}%
\providecommand \href [0]{\begingroup \@sanitize@url \@href}%
\providecommand \@href[1]{\@@startlink{#1}\@@href}%
\providecommand \@@href[1]{\endgroup#1\@@endlink}%
\providecommand \@sanitize@url [0]{\catcode `\\12\catcode `\$12\catcode
  `\&12\catcode `\#12\catcode `\^12\catcode `\_12\catcode `\%12\relax}%
\providecommand \@@startlink[1]{}%
\providecommand \@@endlink[0]{}%
\providecommand \url  [0]{\begingroup\@sanitize@url \@url }%
\providecommand \@url [1]{\endgroup\@href {#1}{\urlprefix }}%
\providecommand \urlprefix  [0]{URL }%
\providecommand \Eprint [0]{\href }%
\providecommand \doibase [0]{http://dx.doi.org/}%
\providecommand \selectlanguage [0]{\@gobble}%
\providecommand \bibinfo  [0]{\@secondoftwo}%
\providecommand \bibfield  [0]{\@secondoftwo}%
\providecommand \translation [1]{[#1]}%
\providecommand \BibitemOpen [0]{}%
\providecommand \bibitemStop [0]{}%
\providecommand \bibitemNoStop [0]{.\EOS\space}%
\providecommand \EOS [0]{\spacefactor3000\relax}%
\providecommand \BibitemShut  [1]{\csname bibitem#1\endcsname}%
\let\auto@bib@innerbib\@empty
\bibitem [{\citenamefont {Wineland}\ \emph {et~al.}(1998)\citenamefont
  {Wineland}, \citenamefont {Monroe}, \citenamefont {Meekhof}, \citenamefont
  {King}, \citenamefont {Leibfried}, \citenamefont {Itano}, \citenamefont
  {Bergquist}, \citenamefont {Berkeland}, \citenamefont {Bollinger},\ and\
  \citenamefont {Miller}}]{Wineland1998}%
  \BibitemOpen
  \bibfield  {author} {\bibinfo {author} {\bibfnamefont {D.~J.}\ \bibnamefont
  {Wineland}}, \bibinfo {author} {\bibfnamefont {C.}~\bibnamefont {Monroe}},
  \bibinfo {author} {\bibfnamefont {D.~M.}\ \bibnamefont {Meekhof}}, \bibinfo
  {author} {\bibfnamefont {B.~E.}\ \bibnamefont {King}}, \bibinfo {author}
  {\bibfnamefont {D.}~\bibnamefont {Leibfried}}, \bibinfo {author}
  {\bibfnamefont {W.}~\bibnamefont {Itano}}, \bibinfo {author} {\bibfnamefont
  {J.~C.}\ \bibnamefont {Bergquist}}, \bibinfo {author} {\bibfnamefont
  {D.}~\bibnamefont {Berkeland}}, \bibinfo {author} {\bibfnamefont {J.~J.}\
  \bibnamefont {Bollinger}}, \ and\ \bibinfo {author} {\bibfnamefont
  {J.}~\bibnamefont {Miller}},\ }\href@noop {} {\bibfield  {journal} {\bibinfo
  {journal} {Proc. R. Soc. A}\ }\textbf {\bibinfo {volume} {454}},\ \bibinfo
  {pages} {411} (\bibinfo {year} {1998})}\BibitemShut {NoStop}%
\bibitem [{\citenamefont {Olmschenk}\ \emph {et~al.}(2009)\citenamefont
  {Olmschenk}, \citenamefont {Matsukevich}, \citenamefont {Maunz},
  \citenamefont {Hayes}, \citenamefont {Duan},\ and\ \citenamefont
  {Monroe}}]{Olmschenk2009}%
  \BibitemOpen
  \bibfield  {author} {\bibinfo {author} {\bibfnamefont {S.}~\bibnamefont
  {Olmschenk}}, \bibinfo {author} {\bibfnamefont {D.~N.}\ \bibnamefont
  {Matsukevich}}, \bibinfo {author} {\bibfnamefont {P.}~\bibnamefont {Maunz}},
  \bibinfo {author} {\bibfnamefont {D.}~\bibnamefont {Hayes}}, \bibinfo
  {author} {\bibfnamefont {L.-M.}\ \bibnamefont {Duan}}, \ and\ \bibinfo
  {author} {\bibfnamefont {C.}~\bibnamefont {Monroe}},\ }\href@noop {}
  {\bibfield  {journal} {\bibinfo  {journal} {Science}\ }\textbf {\bibinfo
  {volume} {323}},\ \bibinfo {pages} {486} (\bibinfo {year}
  {2009})}\BibitemShut {NoStop}%
\bibitem [{\citenamefont {Hofmann}\ \emph {et~al.}(2012)\citenamefont
  {Hofmann}, \citenamefont {Krug}, \citenamefont {Ortegel}, \citenamefont
  {Gerard}, \citenamefont {Weber}, \citenamefont {Rosenfeld},\ and\
  \citenamefont {Weinfurter}}]{Hofmann2012}%
  \BibitemOpen
  \bibfield  {author} {\bibinfo {author} {\bibfnamefont {J.}~\bibnamefont
  {Hofmann}}, \bibinfo {author} {\bibfnamefont {M.}~\bibnamefont {Krug}},
  \bibinfo {author} {\bibfnamefont {N.}~\bibnamefont {Ortegel}}, \bibinfo
  {author} {\bibfnamefont {L.}~\bibnamefont {Gerard}}, \bibinfo {author}
  {\bibfnamefont {M.}~\bibnamefont {Weber}}, \bibinfo {author} {\bibfnamefont
  {W.}~\bibnamefont {Rosenfeld}}, \ and\ \bibinfo {author} {\bibfnamefont
  {H.}~\bibnamefont {Weinfurter}},\ }\href@noop {} {\bibfield  {journal}
  {\bibinfo  {journal} {Science}\ }\textbf {\bibinfo {volume} {337}},\ \bibinfo
  {pages} {72} (\bibinfo {year} {2012})}\BibitemShut {NoStop}%
\bibitem [{\citenamefont {Schellekens}\ \emph {et~al.}(2005)\citenamefont
  {Schellekens}, \citenamefont {Hoppeler}, \citenamefont {Perrin},
  \citenamefont {Gomes}, \citenamefont {Boiron}, \citenamefont {Aspect},\ and\
  \citenamefont {Westbrook}}]{Schellekens2005}%
  \BibitemOpen
  \bibfield  {author} {\bibinfo {author} {\bibfnamefont {M.}~\bibnamefont
  {Schellekens}}, \bibinfo {author} {\bibfnamefont {R.}~\bibnamefont
  {Hoppeler}}, \bibinfo {author} {\bibfnamefont {A.}~\bibnamefont {Perrin}},
  \bibinfo {author} {\bibfnamefont {J.~V.}\ \bibnamefont {Gomes}}, \bibinfo
  {author} {\bibfnamefont {D.}~\bibnamefont {Boiron}}, \bibinfo {author}
  {\bibfnamefont {A.}~\bibnamefont {Aspect}}, \ and\ \bibinfo {author}
  {\bibfnamefont {C.~I.}\ \bibnamefont {Westbrook}},\ }\href@noop {} {\bibfield
   {journal} {\bibinfo  {journal} {Science}\ }\textbf {\bibinfo {volume}
  {310}},\ \bibinfo {pages} {648} (\bibinfo {year} {2005})}\BibitemShut
  {NoStop}%
\bibitem [{\citenamefont {Beugnon}\ \emph {et~al.}(2006)\citenamefont
  {Beugnon}, \citenamefont {Jones}, \citenamefont {Dingjan}, \citenamefont
  {Darqui{\'e}}, \citenamefont {Messin}, \citenamefont {Browaeys},\ and\
  \citenamefont {Grangier}}]{Beugnon2006}%
  \BibitemOpen
  \bibfield  {author} {\bibinfo {author} {\bibfnamefont {J.}~\bibnamefont
  {Beugnon}}, \bibinfo {author} {\bibfnamefont {M.~P.~A.}\ \bibnamefont
  {Jones}}, \bibinfo {author} {\bibfnamefont {J.}~\bibnamefont {Dingjan}},
  \bibinfo {author} {\bibfnamefont {B.}~\bibnamefont {Darqui{\'e}}}, \bibinfo
  {author} {\bibfnamefont {G.}~\bibnamefont {Messin}}, \bibinfo {author}
  {\bibfnamefont {A.}~\bibnamefont {Browaeys}}, \ and\ \bibinfo {author}
  {\bibfnamefont {P.}~\bibnamefont {Grangier}},\ }\href@noop {} {\bibfield
  {journal} {\bibinfo  {journal} {Nature (London)}\ }\textbf {\bibinfo {volume}
  {440}},\ \bibinfo {pages} {779} (\bibinfo {year} {2006})}\BibitemShut
  {NoStop}%
\bibitem [{\citenamefont {Gericke}\ \emph {et~al.}(2008)\citenamefont
  {Gericke}, \citenamefont {W\"{u}rtz}, \citenamefont {Reitz}, \citenamefont
  {Langen},\ and\ \citenamefont {Ott}}]{Gericke2008}%
  \BibitemOpen
  \bibfield  {author} {\bibinfo {author} {\bibfnamefont {T.}~\bibnamefont
  {Gericke}}, \bibinfo {author} {\bibfnamefont {P.}~\bibnamefont {W\"{u}rtz}},
  \bibinfo {author} {\bibfnamefont {D.}~\bibnamefont {Reitz}}, \bibinfo
  {author} {\bibfnamefont {T.}~\bibnamefont {Langen}}, \ and\ \bibinfo {author}
  {\bibfnamefont {H.}~\bibnamefont {Ott}},\ }\href@noop {} {\bibfield
  {journal} {\bibinfo  {journal} {Nat. Phys.}\ }\textbf {\bibinfo {volume}
  {4}},\ \bibinfo {pages} {949} (\bibinfo {year} {2008})}\BibitemShut {NoStop}%
\bibitem [{\citenamefont {Bakr}\ \emph {et~al.}(2009)\citenamefont {Bakr},
  \citenamefont {Gillen}, \citenamefont {Peng}, \citenamefont {F\"{o}lling},\
  and\ \citenamefont {Greiner}}]{Bakr2009}%
  \BibitemOpen
  \bibfield  {author} {\bibinfo {author} {\bibfnamefont {W.~S.}\ \bibnamefont
  {Bakr}}, \bibinfo {author} {\bibfnamefont {J.~I.}\ \bibnamefont {Gillen}},
  \bibinfo {author} {\bibfnamefont {A.}~\bibnamefont {Peng}}, \bibinfo {author}
  {\bibfnamefont {S.}~\bibnamefont {F\"{o}lling}}, \ and\ \bibinfo {author}
  {\bibfnamefont {M.}~\bibnamefont {Greiner}},\ }\href@noop {} {\bibfield
  {journal} {\bibinfo  {journal} {Nature (London)}\ }\textbf {\bibinfo {volume}
  {462}},\ \bibinfo {pages} {74} (\bibinfo {year} {2009})}\BibitemShut
  {NoStop}%
\bibitem [{\citenamefont {Weitenberg}\ \emph {et~al.}(2011)\citenamefont
  {Weitenberg}, \citenamefont {Endres}, \citenamefont {Sherson}, \citenamefont
  {Cheneau}, \citenamefont {Schau{\ss}}, \citenamefont {Fukuhara},
  \citenamefont {Bloch},\ and\ \citenamefont {Kuhr}}]{Weitenberg2011}%
  \BibitemOpen
  \bibfield  {author} {\bibinfo {author} {\bibfnamefont {C.}~\bibnamefont
  {Weitenberg}}, \bibinfo {author} {\bibfnamefont {M.}~\bibnamefont {Endres}},
  \bibinfo {author} {\bibfnamefont {J.~F.}\ \bibnamefont {Sherson}}, \bibinfo
  {author} {\bibfnamefont {M.}~\bibnamefont {Cheneau}}, \bibinfo {author}
  {\bibfnamefont {P.}~\bibnamefont {Schau{\ss}}}, \bibinfo {author}
  {\bibfnamefont {T.}~\bibnamefont {Fukuhara}}, \bibinfo {author}
  {\bibfnamefont {I.}~\bibnamefont {Bloch}}, \ and\ \bibinfo {author}
  {\bibfnamefont {S.}~\bibnamefont {Kuhr}},\ }\href@noop {} {\bibfield
  {journal} {\bibinfo  {journal} {Nature (London)}\ }\textbf {\bibinfo {volume}
  {471}},\ \bibinfo {pages} {319} (\bibinfo {year} {2011})}\BibitemShut
  {NoStop}%
\bibitem [{\citenamefont {Fuhrmanek}\ \emph {et~al.}(2011)\citenamefont
  {Fuhrmanek}, \citenamefont {Bourgain}, \citenamefont {Sortais},\ and\
  \citenamefont {Browaeys}}]{Fuhrmanek2011}%
  \BibitemOpen
  \bibfield  {author} {\bibinfo {author} {\bibfnamefont {A.}~\bibnamefont
  {Fuhrmanek}}, \bibinfo {author} {\bibfnamefont {R.}~\bibnamefont {Bourgain}},
  \bibinfo {author} {\bibfnamefont {Y.~R.~P.}\ \bibnamefont {Sortais}}, \ and\
  \bibinfo {author} {\bibfnamefont {A.}~\bibnamefont {Browaeys}},\ }\href@noop
  {} {\bibfield  {journal} {\bibinfo  {journal} {Phys. Rev. Lett.}\ }\textbf
  {\bibinfo {volume} {106}},\ \bibinfo {pages} {133003} (\bibinfo {year}
  {2011})}\BibitemShut {NoStop}%
\bibitem [{\citenamefont {Gibbons}\ \emph {et~al.}(2011)\citenamefont
  {Gibbons}, \citenamefont {Hamley}, \citenamefont {Shih},\ and\ \citenamefont
  {Chapman}}]{Gibbons2011}%
  \BibitemOpen
  \bibfield  {author} {\bibinfo {author} {\bibfnamefont {M.~J.}\ \bibnamefont
  {Gibbons}}, \bibinfo {author} {\bibfnamefont {C.~D.}\ \bibnamefont {Hamley}},
  \bibinfo {author} {\bibfnamefont {C.-Y.}\ \bibnamefont {Shih}}, \ and\
  \bibinfo {author} {\bibfnamefont {M.~S.}\ \bibnamefont {Chapman}},\
  }\href@noop {} {\bibfield  {journal} {\bibinfo  {journal} {Phys. Rev. Lett.}\
  }\textbf {\bibinfo {volume} {106}},\ \bibinfo {pages} {133002} (\bibinfo
  {year} {2011})}\BibitemShut {NoStop}%
\bibitem [{\citenamefont {Schlosser}\ \emph {et~al.}(2001)\citenamefont
  {Schlosser}, \citenamefont {Reymond}, \citenamefont {Protsenko},\ and\
  \citenamefont {Grangier}}]{Schlosser2001}%
  \BibitemOpen
  \bibfield  {author} {\bibinfo {author} {\bibfnamefont {N.}~\bibnamefont
  {Schlosser}}, \bibinfo {author} {\bibfnamefont {G.}~\bibnamefont {Reymond}},
  \bibinfo {author} {\bibfnamefont {I.}~\bibnamefont {Protsenko}}, \ and\
  \bibinfo {author} {\bibfnamefont {P.}~\bibnamefont {Grangier}},\ }\href@noop
  {} {\bibfield  {journal} {\bibinfo  {journal} {Nature (London)}\ }\textbf
  {\bibinfo {volume} {411}},\ \bibinfo {pages} {1024} (\bibinfo {year}
  {2001})}\BibitemShut {NoStop}%
\bibitem [{\citenamefont {Sherson}\ \emph {et~al.}(2010)\citenamefont
  {Sherson}, \citenamefont {Weitenberg}, \citenamefont {Endres}, \citenamefont
  {Cheneau}, \citenamefont {Bloch},\ and\ \citenamefont {Kuhr}}]{Sherson2010}%
  \BibitemOpen
  \bibfield  {author} {\bibinfo {author} {\bibfnamefont {J.}~\bibnamefont
  {Sherson}}, \bibinfo {author} {\bibfnamefont {C.}~\bibnamefont {Weitenberg}},
  \bibinfo {author} {\bibfnamefont {M.}~\bibnamefont {Endres}}, \bibinfo
  {author} {\bibfnamefont {M.}~\bibnamefont {Cheneau}}, \bibinfo {author}
  {\bibfnamefont {I.}~\bibnamefont {Bloch}}, \ and\ \bibinfo {author}
  {\bibfnamefont {S.}~\bibnamefont {Kuhr}},\ }\href@noop {} {\bibfield
  {journal} {\bibinfo  {journal} {Nature (London)}\ }\textbf {\bibinfo {volume}
  {467}},\ \bibinfo {pages} {68} (\bibinfo {year} {2010})}\BibitemShut
  {NoStop}%
\bibitem [{\citenamefont {Nelson}\ \emph {et~al.}(2007)\citenamefont {Nelson},
  \citenamefont {Li},\ and\ \citenamefont {Weiss}}]{Nelson2007}%
  \BibitemOpen
  \bibfield  {author} {\bibinfo {author} {\bibfnamefont {K.~D.}\ \bibnamefont
  {Nelson}}, \bibinfo {author} {\bibfnamefont {X.}~\bibnamefont {Li}}, \ and\
  \bibinfo {author} {\bibfnamefont {D.~S.}\ \bibnamefont {Weiss}},\ }\href@noop
  {} {\bibfield  {journal} {\bibinfo  {journal} {Nat. Phys.}\ }\textbf
  {\bibinfo {volume} {3}},\ \bibinfo {pages} {556} (\bibinfo {year}
  {2007})}\BibitemShut {NoStop}%
\bibitem [{\citenamefont {Setiawan}(2012)}]{Setiawan2012thesis}%
  \BibitemOpen
  \bibfield  {author} {\bibinfo {author} {\bibfnamefont {W.}~\bibnamefont
  {Setiawan}},\ }\emph {\bibinfo {title} {{Fermi Gas Microscope}}},\ \href@noop
  {} {Ph.D. thesis},\ \bibinfo  {school} {Harvard University} (\bibinfo {year}
  {2012})\BibitemShut {NoStop}%
\bibitem [{\citenamefont {Landini}\ \emph {et~al.}(2011)\citenamefont
  {Landini}, \citenamefont {Roy}, \citenamefont {Carcagn\'{i}}, \citenamefont
  {Trypogeorgos}, \citenamefont {Fattori}, \citenamefont {Inguscio},\ and\
  \citenamefont {Modugno}}]{Landini2011}%
  \BibitemOpen
  \bibfield  {author} {\bibinfo {author} {\bibfnamefont {M.}~\bibnamefont
  {Landini}}, \bibinfo {author} {\bibfnamefont {S.}~\bibnamefont {Roy}},
  \bibinfo {author} {\bibfnamefont {L.}~\bibnamefont {Carcagn\'{i}}}, \bibinfo
  {author} {\bibfnamefont {D.}~\bibnamefont {Trypogeorgos}}, \bibinfo {author}
  {\bibfnamefont {M.}~\bibnamefont {Fattori}}, \bibinfo {author} {\bibfnamefont
  {M.}~\bibnamefont {Inguscio}}, \ and\ \bibinfo {author} {\bibfnamefont
  {G.}~\bibnamefont {Modugno}},\ }\href@noop {} {\bibfield  {journal} {\bibinfo
   {journal} {Phys. Rev. A}\ }\textbf {\bibinfo {volume} {84}},\ \bibinfo
  {pages} {043432} (\bibinfo {year} {2011})}\BibitemShut {NoStop}%
\bibitem [{\citenamefont {Fernandes}\ \emph {et~al.}(2012)\citenamefont
  {Fernandes}, \citenamefont {Sievers}, \citenamefont {Kretzschmar},
  \citenamefont {Wu}, \citenamefont {Salomon},\ and\ \citenamefont
  {Chevy}}]{Fernandes2012}%
  \BibitemOpen
  \bibfield  {author} {\bibinfo {author} {\bibfnamefont {D.~R.}\ \bibnamefont
  {Fernandes}}, \bibinfo {author} {\bibfnamefont {F.}~\bibnamefont {Sievers}},
  \bibinfo {author} {\bibfnamefont {N.}~\bibnamefont {Kretzschmar}}, \bibinfo
  {author} {\bibfnamefont {S.}~\bibnamefont {Wu}}, \bibinfo {author}
  {\bibfnamefont {C.}~\bibnamefont {Salomon}}, \ and\ \bibinfo {author}
  {\bibfnamefont {F.}~\bibnamefont {Chevy}},\ }\href@noop {} {\bibfield
  {journal} {\bibinfo  {journal} {Europhys. Lett.}\ }\textbf {\bibinfo {volume}
  {100}},\ \bibinfo {pages} {63001} (\bibinfo {year} {2012})}\BibitemShut
  {NoStop}%
\bibitem [{\citenamefont {Z\"{u}rn}\ \emph {et~al.}(2012)\citenamefont
  {Z\"{u}rn}, \citenamefont {Serwane}, \citenamefont {Lompe}, \citenamefont
  {Wenz}, \citenamefont {Ries}, \citenamefont {Bohn},\ and\ \citenamefont
  {Jochim}}]{Zurn2012}%
  \BibitemOpen
  \bibfield  {author} {\bibinfo {author} {\bibfnamefont {G.}~\bibnamefont
  {Z\"{u}rn}}, \bibinfo {author} {\bibfnamefont {F.}~\bibnamefont {Serwane}},
  \bibinfo {author} {\bibfnamefont {T.}~\bibnamefont {Lompe}}, \bibinfo
  {author} {\bibfnamefont {A.~N.}\ \bibnamefont {Wenz}}, \bibinfo {author}
  {\bibfnamefont {M.~G.}\ \bibnamefont {Ries}}, \bibinfo {author}
  {\bibfnamefont {J.~E.}\ \bibnamefont {Bohn}}, \ and\ \bibinfo {author}
  {\bibfnamefont {S.}~\bibnamefont {Jochim}},\ }\href@noop {} {\bibfield
  {journal} {\bibinfo  {journal} {Phys. Rev. Lett.}\ }\textbf {\bibinfo
  {volume} {108}},\ \bibinfo {pages} {075303} (\bibinfo {year}
  {2012})}\BibitemShut {NoStop}%
\bibitem [{\citenamefont {Yefsah}\ \emph {et~al.}(2013)\citenamefont {Yefsah},
  \citenamefont {Sommer}, \citenamefont {Ku}, \citenamefont {Cheuk},
  \citenamefont {Ji}, \citenamefont {Bakr},\ and\ \citenamefont
  {Zwierlein}}]{Yefsah2013}%
  \BibitemOpen
  \bibfield  {author} {\bibinfo {author} {\bibfnamefont {T.}~\bibnamefont
  {Yefsah}}, \bibinfo {author} {\bibfnamefont {A.~T.}\ \bibnamefont {Sommer}},
  \bibinfo {author} {\bibfnamefont {M.~J.~H.}\ \bibnamefont {Ku}}, \bibinfo
  {author} {\bibfnamefont {L.~W.}\ \bibnamefont {Cheuk}}, \bibinfo {author}
  {\bibfnamefont {W.}~\bibnamefont {Ji}}, \bibinfo {author} {\bibfnamefont
  {W.~S.}\ \bibnamefont {Bakr}}, \ and\ \bibinfo {author} {\bibfnamefont
  {M.~W.}\ \bibnamefont {Zwierlein}},\ }\href@noop {} {\bibfield  {journal}
  {\bibinfo  {journal} {Nature (London)}\ }\textbf {\bibinfo {volume} {499}},\
  \bibinfo {pages} {426} (\bibinfo {year} {2013})}\BibitemShut {NoStop}%
\bibitem [{\citenamefont {Monroe}\ \emph {et~al.}(1995)\citenamefont {Monroe},
  \citenamefont {Meekhof}, \citenamefont {King}, \citenamefont {Jefferts},
  \citenamefont {Itano}, \citenamefont {Wineland},\ and\ \citenamefont
  {Gould}}]{Monroe1995}%
  \BibitemOpen
  \bibfield  {author} {\bibinfo {author} {\bibfnamefont {C.}~\bibnamefont
  {Monroe}}, \bibinfo {author} {\bibfnamefont {D.~M.}\ \bibnamefont {Meekhof}},
  \bibinfo {author} {\bibfnamefont {B.~E.}\ \bibnamefont {King}}, \bibinfo
  {author} {\bibfnamefont {S.~R.}\ \bibnamefont {Jefferts}}, \bibinfo {author}
  {\bibfnamefont {W.~M.}\ \bibnamefont {Itano}}, \bibinfo {author}
  {\bibfnamefont {D.~J.}\ \bibnamefont {Wineland}}, \ and\ \bibinfo {author}
  {\bibfnamefont {P.}~\bibnamefont {Gould}},\ }\href@noop {} {\bibfield
  {journal} {\bibinfo  {journal} {Phys. Rev. Lett.}\ }\textbf {\bibinfo
  {volume} {75}},\ \bibinfo {pages} {4011} (\bibinfo {year}
  {1995})}\BibitemShut {NoStop}%
\bibitem [{\citenamefont {Kerman}\ \emph {et~al.}(2000)\citenamefont {Kerman},
  \citenamefont {Vuletic}, \citenamefont {Chin},\ and\ \citenamefont
  {Chu}}]{Kerman2000}%
  \BibitemOpen
  \bibfield  {author} {\bibinfo {author} {\bibfnamefont {A.~J.}\ \bibnamefont
  {Kerman}}, \bibinfo {author} {\bibfnamefont {V.~V.}\ \bibnamefont {Vuletic}},
  \bibinfo {author} {\bibfnamefont {C.}~\bibnamefont {Chin}}, \ and\ \bibinfo
  {author} {\bibfnamefont {S.}~\bibnamefont {Chu}},\ }\href@noop {} {\bibfield
  {journal} {\bibinfo  {journal} {Phys. Rev. Lett.}\ }\textbf {\bibinfo
  {volume} {84}},\ \bibinfo {pages} {439} (\bibinfo {year} {2000})}\BibitemShut
  {NoStop}%
\bibitem [{\citenamefont {Han}\ \emph {et~al.}(2000)\citenamefont {Han},
  \citenamefont {Wolf}, \citenamefont {Oliver}, \citenamefont {McCormick},
  \citenamefont {DePue},\ and\ \citenamefont {Weiss}}]{Han2000}%
  \BibitemOpen
  \bibfield  {author} {\bibinfo {author} {\bibfnamefont {D.-J.}\ \bibnamefont
  {Han}}, \bibinfo {author} {\bibfnamefont {S.}~\bibnamefont {Wolf}}, \bibinfo
  {author} {\bibfnamefont {S.}~\bibnamefont {Oliver}}, \bibinfo {author}
  {\bibfnamefont {C.}~\bibnamefont {McCormick}}, \bibinfo {author}
  {\bibfnamefont {M.~T.}\ \bibnamefont {DePue}}, \ and\ \bibinfo {author}
  {\bibfnamefont {D.~S.}\ \bibnamefont {Weiss}},\ }\href@noop {} {\bibfield
  {journal} {\bibinfo  {journal} {Phys. Rev. Lett.}\ }\textbf {\bibinfo
  {volume} {85}},\ \bibinfo {pages} {724} (\bibinfo {year} {2000})}\BibitemShut
  {NoStop}%
\bibitem [{\citenamefont {Li}\ \emph {et~al.}(2012)\citenamefont {Li},
  \citenamefont {Corcovilos}, \citenamefont {Wang},\ and\ \citenamefont
  {Weiss}}]{Li2012}%
  \BibitemOpen
  \bibfield  {author} {\bibinfo {author} {\bibfnamefont {X.}~\bibnamefont
  {Li}}, \bibinfo {author} {\bibfnamefont {T.~A.}\ \bibnamefont {Corcovilos}},
  \bibinfo {author} {\bibfnamefont {Y.}~\bibnamefont {Wang}}, \ and\ \bibinfo
  {author} {\bibfnamefont {D.~S.}\ \bibnamefont {Weiss}},\ }\href@noop {}
  {\bibfield  {journal} {\bibinfo  {journal} {Phys. Rev. Lett.}\ }\textbf
  {\bibinfo {volume} {108}},\ \bibinfo {pages} {103001} (\bibinfo {year}
  {2012})}\BibitemShut {NoStop}%
\bibitem [{\citenamefont {Kaufman}\ \emph {et~al.}(2012)\citenamefont
  {Kaufman}, \citenamefont {Lester},\ and\ \citenamefont
  {Regal}}]{Kaufman2012}%
  \BibitemOpen
  \bibfield  {author} {\bibinfo {author} {\bibfnamefont {A.~M.}\ \bibnamefont
  {Kaufman}}, \bibinfo {author} {\bibfnamefont {B.~J.}\ \bibnamefont {Lester}},
  \ and\ \bibinfo {author} {\bibfnamefont {C.~A.}\ \bibnamefont {Regal}},\
  }\href@noop {} {\bibfield  {journal} {\bibinfo  {journal} {Phys. Rev. X}\
  }\textbf {\bibinfo {volume} {2}},\ \bibinfo {pages} {041014} (\bibinfo {year}
  {2012})}\BibitemShut {NoStop}%
\bibitem [{\citenamefont {Diedrich}\ \emph {et~al.}(1989)\citenamefont
  {Diedrich}, \citenamefont {Bergquist}, \citenamefont {Itano},\ and\
  \citenamefont {Wineland}}]{Diedrich1989}%
  \BibitemOpen
  \bibfield  {author} {\bibinfo {author} {\bibfnamefont {F.}~\bibnamefont
  {Diedrich}}, \bibinfo {author} {\bibfnamefont {J.~C.}\ \bibnamefont
  {Bergquist}}, \bibinfo {author} {\bibfnamefont {W.~M.}\ \bibnamefont
  {Itano}}, \ and\ \bibinfo {author} {\bibfnamefont {D.~J.}\ \bibnamefont
  {Wineland}},\ }\href@noop {} {\bibfield  {journal} {\bibinfo  {journal}
  {Phys. Rev. Lett.}\ }\textbf {\bibinfo {volume} {62}},\ \bibinfo {pages}
  {403} (\bibinfo {year} {1989})}\BibitemShut {NoStop}%
\bibitem [{\citenamefont {Hamann}\ \emph {et~al.}(1998)\citenamefont {Hamann},
  \citenamefont {Haycock}, \citenamefont {Klose}, \citenamefont {Pax},
  \citenamefont {Deutsch},\ and\ \citenamefont {Jessen}}]{Hamann1998}%
  \BibitemOpen
  \bibfield  {author} {\bibinfo {author} {\bibfnamefont {S.~E.}\ \bibnamefont
  {Hamann}}, \bibinfo {author} {\bibfnamefont {D.~L.}\ \bibnamefont {Haycock}},
  \bibinfo {author} {\bibfnamefont {G.}~\bibnamefont {Klose}}, \bibinfo
  {author} {\bibfnamefont {P.~H.}\ \bibnamefont {Pax}}, \bibinfo {author}
  {\bibfnamefont {I.~H.}\ \bibnamefont {Deutsch}}, \ and\ \bibinfo {author}
  {\bibfnamefont {P.~S.}\ \bibnamefont {Jessen}},\ }\href@noop {} {\bibfield
  {journal} {\bibinfo  {journal} {Phys. Rev. Lett.}\ }\textbf {\bibinfo
  {volume} {80}},\ \bibinfo {pages} {4149} (\bibinfo {year}
  {1998})}\BibitemShut {NoStop}%
\bibitem [{\citenamefont {Perrin}\ \emph {et~al.}(1998)\citenamefont {Perrin},
  \citenamefont {Kuhn}, \citenamefont {Bouchoule},\ and\ \citenamefont
  {Salomon}}]{Perrin1998}%
  \BibitemOpen
  \bibfield  {author} {\bibinfo {author} {\bibfnamefont {H.}~\bibnamefont
  {Perrin}}, \bibinfo {author} {\bibfnamefont {A.}~\bibnamefont {Kuhn}},
  \bibinfo {author} {\bibfnamefont {I.}~\bibnamefont {Bouchoule}}, \ and\
  \bibinfo {author} {\bibfnamefont {C.}~\bibnamefont {Salomon}},\ }\href@noop
  {} {\bibfield  {journal} {\bibinfo  {journal} {Europhys. Lett.}\ }\textbf
  {\bibinfo {volume} {42}},\ \bibinfo {pages} {395} (\bibinfo {year}
  {1998})}\BibitemShut {NoStop}%
\bibitem [{\citenamefont {Forster}\ \emph {et~al.}(2009)\citenamefont
  {Forster}, \citenamefont {Karski}, \citenamefont {Choi}, \citenamefont
  {Steffen}, \citenamefont {Alt}, \citenamefont {Meschede}, \citenamefont
  {Widera}, \citenamefont {Montano}, \citenamefont {Lee}, \citenamefont
  {Rakreungdet},\ and\ \citenamefont {Jessen}}]{Forster2009}%
  \BibitemOpen
  \bibfield  {author} {\bibinfo {author} {\bibfnamefont {L.}~\bibnamefont
  {Forster}}, \bibinfo {author} {\bibfnamefont {M.}~\bibnamefont {Karski}},
  \bibinfo {author} {\bibfnamefont {J.-M.}\ \bibnamefont {Choi}}, \bibinfo
  {author} {\bibfnamefont {A.}~\bibnamefont {Steffen}}, \bibinfo {author}
  {\bibfnamefont {W.}~\bibnamefont {Alt}}, \bibinfo {author} {\bibfnamefont
  {D.}~\bibnamefont {Meschede}}, \bibinfo {author} {\bibfnamefont
  {A.}~\bibnamefont {Widera}}, \bibinfo {author} {\bibfnamefont
  {E.}~\bibnamefont {Montano}}, \bibinfo {author} {\bibfnamefont {J.~H.}\
  \bibnamefont {Lee}}, \bibinfo {author} {\bibfnamefont {W.}~\bibnamefont
  {Rakreungdet}}, \ and\ \bibinfo {author} {\bibfnamefont {P.~S.}\ \bibnamefont
  {Jessen}},\ }\href@noop {} {\bibfield  {journal} {\bibinfo  {journal} {Phys.
  Rev. Lett.}\ }\textbf {\bibinfo {volume} {103}},\ \bibinfo {pages} {233001}
  (\bibinfo {year} {2009})}\BibitemShut {NoStop}%
\bibitem [{\citenamefont {Kaufman}\ \emph {et~al.}(0057)\citenamefont
  {Kaufman}, \citenamefont {Lester}, \citenamefont {Reynolds}, \citenamefont
  {Wall}, \citenamefont {Foss-Feig}, \citenamefont {Hazzard}, \citenamefont
  {Rey},\ and\ \citenamefont {Regal}}]{Kaufman2014}%
  \BibitemOpen
  \bibfield  {author} {\bibinfo {author} {\bibfnamefont {A.~M.}\ \bibnamefont
  {Kaufman}}, \bibinfo {author} {\bibfnamefont {B.~J.}\ \bibnamefont {Lester}},
  \bibinfo {author} {\bibfnamefont {C.~M.}\ \bibnamefont {Reynolds}}, \bibinfo
  {author} {\bibfnamefont {M.~L.}\ \bibnamefont {Wall}}, \bibinfo {author}
  {\bibfnamefont {M.}~\bibnamefont {Foss-Feig}}, \bibinfo {author}
  {\bibfnamefont {K.~R.~A.}\ \bibnamefont {Hazzard}}, \bibinfo {author}
  {\bibfnamefont {A.~M.}\ \bibnamefont {Rey}}, \ and\ \bibinfo {author}
  {\bibfnamefont {C.~A.}\ \bibnamefont {Regal}},\ }\href@noop {} {\bibfield
  {journal}  {\bibinfo  {journal} {Science}\
  }\textbf {\bibinfo {volume} {345}},\ \bibinfo {pages} {306} (\bibinfo
  {year} {2014})}\BibitemShut {NoStop}%
\bibitem [{\citenamefont {Schlosser}\ \emph {et~al.}(2002)\citenamefont
  {Schlosser}, \citenamefont {Reymond},\ and\ \citenamefont
  {Grangier}}]{Schlosser2002}%
  \BibitemOpen
  \bibfield  {author} {\bibinfo {author} {\bibfnamefont {N.}~\bibnamefont
  {Schlosser}}, \bibinfo {author} {\bibfnamefont {G.}~\bibnamefont {Reymond}},
  \ and\ \bibinfo {author} {\bibfnamefont {P.}~\bibnamefont {Grangier}},\
  }\href@noop {} {\bibfield  {journal} {\bibinfo  {journal} {Phys. Rev. Lett.}\
  }\textbf {\bibinfo {volume} {89}},\ \bibinfo {pages} {023005} (\bibinfo
  {year} {2002})}\BibitemShut {NoStop}%
\bibitem [{\citenamefont {Sompet}\ \emph {et~al.}(2013)\citenamefont {Sompet},
  \citenamefont {Carpentier}, \citenamefont {Fung}, \citenamefont {McGovern},\
  and\ \citenamefont {Andersen}}]{Sompet2013}%
  \BibitemOpen
  \bibfield  {author} {\bibinfo {author} {\bibfnamefont {P.}~\bibnamefont
  {Sompet}}, \bibinfo {author} {\bibfnamefont {A.~V.}\ \bibnamefont
  {Carpentier}}, \bibinfo {author} {\bibfnamefont {Y.~H.}\ \bibnamefont
  {Fung}}, \bibinfo {author} {\bibfnamefont {M.}~\bibnamefont {McGovern}}, \
  and\ \bibinfo {author} {\bibfnamefont {M.~F.}\ \bibnamefont {Andersen}},\
  }\href@noop {} {\bibfield  {journal} {\bibinfo  {journal} {Phys. Rev. A}\
  }\textbf {\bibinfo {volume} {88}},\ \bibinfo {pages} {051401(R)} (\bibinfo
  {year} {2013})}\BibitemShut {NoStop}%
\bibitem [{\citenamefont {Dalibard}\ and\ \citenamefont
  {Cohen-Tannoudji}(1989)}]{Dalibard1989}%
  \BibitemOpen
  \bibfield  {author} {\bibinfo {author} {\bibfnamefont {J.}~\bibnamefont
  {Dalibard}}\ and\ \bibinfo {author} {\bibfnamefont {C.}~\bibnamefont
  {Cohen-Tannoudji}},\ }\href@noop {} {\bibfield  {journal} {\bibinfo
  {journal} {J. Opt. Soc. Am. B}\ }\textbf {\bibinfo {volume} {6}},\ \bibinfo
  {pages} {2023} (\bibinfo {year} {1989})}\BibitemShut {NoStop}%
\bibitem [{\citenamefont {Tuchendler}\ \emph {et~al.}(2008)\citenamefont
  {Tuchendler}, \citenamefont {Lance}, \citenamefont {Browaeys}, \citenamefont
  {Sortais},\ and\ \citenamefont {Grangier}}]{Tuchendler2008}%
  \BibitemOpen
  \bibfield  {author} {\bibinfo {author} {\bibfnamefont {C.}~\bibnamefont
  {Tuchendler}}, \bibinfo {author} {\bibfnamefont {A.~M.}\ \bibnamefont
  {Lance}}, \bibinfo {author} {\bibfnamefont {A.}~\bibnamefont {Browaeys}},
  \bibinfo {author} {\bibfnamefont {Y.~R.~P.}\ \bibnamefont {Sortais}}, \ and\
  \bibinfo {author} {\bibfnamefont {P.}~\bibnamefont {Grangier}},\ }\href@noop
  {} {\bibfield  {journal} {\bibinfo  {journal} {Phys. Rev. A}\ }\textbf
  {\bibinfo {volume} {78}},\ \bibinfo {pages} {033425} (\bibinfo {year}
  {2008})}\BibitemShut {NoStop}%
\bibitem [{Note1()}]{Note1}%
  \BibitemOpen
  \bibinfo {note} {Note that the pre- and post-selection images used a six-beam
  PGC configuration, which we find achieves a comparable SNR.}\BibitemShut
  {Stop}%
\bibitem [{\citenamefont {{Le Kein}}\ \emph {et~al.}(2013)\citenamefont {{Le
  Kein}}, \citenamefont {Schneeweiss},\ and\ \citenamefont
  {Rauschenbeutel}}]{LeKein2013}%
  \BibitemOpen
  \bibfield  {author} {\bibinfo {author} {\bibfnamefont {F.}~\bibnamefont {{Le
  Kein}}}, \bibinfo {author} {\bibfnamefont {P.}~\bibnamefont {Schneeweiss}}, \
  and\ \bibinfo {author} {\bibfnamefont {A.}~\bibnamefont {Rauschenbeutel}},\
  }\href@noop {} {\bibfield  {journal} {\bibinfo  {journal} {Eur. Phys. J. D}\
  }\textbf {\bibinfo {volume} {67}},\ \bibinfo {pages} {92} (\bibinfo {year}
  {2013})}\BibitemShut {NoStop}%
\bibitem [{\citenamefont {Safronova}\ and\ \citenamefont
  {Safronova}(2011)}]{Safronova2011}%
  \BibitemOpen
  \bibfield  {author} {\bibinfo {author} {\bibfnamefont {M.~S.}\ \bibnamefont
  {Safronova}}\ and\ \bibinfo {author} {\bibfnamefont {U.~I.}\ \bibnamefont
  {Safronova}},\ }\href@noop {} {\bibfield  {journal} {\bibinfo  {journal}
  {Phys. Rev. A}\ }\textbf {\bibinfo {volume} {83}},\ \bibinfo {pages} {052508}
  (\bibinfo {year} {2011})}\BibitemShut {NoStop}%
\bibitem [{\citenamefont {Arora}\ and\ \citenamefont
  {Sahoo}(2012)}]{Arora2012}%
  \BibitemOpen
  \bibfield  {author} {\bibinfo {author} {\bibfnamefont {B.}~\bibnamefont
  {Arora}}\ and\ \bibinfo {author} {\bibfnamefont {B.~K.}\ \bibnamefont
  {Sahoo}},\ }\href@noop {} {\bibfield  {journal} {\bibinfo  {journal} {Phys.
  Rev. A}\ }\textbf {\bibinfo {volume} {86}},\ \bibinfo {pages} {033416}
  (\bibinfo {year} {2012})}\BibitemShut {NoStop}%
\bibitem [{\citenamefont {Shih}\ and\ \citenamefont
  {Chapman}(2013)}]{Shih2013}%
  \BibitemOpen
  \bibfield  {author} {\bibinfo {author} {\bibfnamefont {C.-Y.}\ \bibnamefont
  {Shih}}\ and\ \bibinfo {author} {\bibfnamefont {M.~S.}\ \bibnamefont
  {Chapman}},\ }\href@noop {} {\bibfield  {journal} {\bibinfo  {journal} {Phys.
  Rev. A}\ }\textbf {\bibinfo {volume} {87}},\ \bibinfo {pages} {063408}
  (\bibinfo {year} {2013})}\BibitemShut {NoStop}%
\bibitem [{\citenamefont {Steck}()}]{Steck2010}%
  \BibitemOpen
  \bibfield  {author} {\bibinfo {author} {\bibfnamefont {D.~A.}\ \bibnamefont
  {Steck}},\ }\href@noop {} {\enquote {\bibinfo {title} {{Rubidium 87 D Line
  Data}},}\ }\bibinfo {howpublished} {available online at
  \url{http://steck.us/alkalidata} (revision 2.1.4, 23 December
  2010)}\BibitemShut {NoStop}%
\bibitem [{Note2()}]{Note2}%
  \BibitemOpen
  \bibinfo {note} {The minimum detection error rate $\delta $ is half the
  overlap integral of the two normalized gaussian distributions, which requires
  an optimal choice of threshold to achieve. The fidelity of the measurement is
  then defined to be $\protect \mathcal {F}=1-\delta $ (i.e., $\delta =\epsilon
  _B=\epsilon _D$ in Ref. \cite {Fuhrmanek2011}).}\BibitemShut {Stop}%
\bibitem [{\citenamefont {Wolf}\ \emph {et~al.}(2000)\citenamefont {Wolf},
  \citenamefont {Oliver},\ and\ \citenamefont {Weiss}}]{Wolf2000}%
  \BibitemOpen
  \bibfield  {author} {\bibinfo {author} {\bibfnamefont {S.}~\bibnamefont
  {Wolf}}, \bibinfo {author} {\bibfnamefont {S.~J.}\ \bibnamefont {Oliver}}, \
  and\ \bibinfo {author} {\bibfnamefont {D.~S.}\ \bibnamefont {Weiss}},\
  }\href@noop {} {\bibfield  {journal} {\bibinfo  {journal} {Phys. Rev. Lett.}\
  }\textbf {\bibinfo {volume} {85}},\ \bibinfo {pages} {4249} (\bibinfo {year}
  {2000})}\BibitemShut {NoStop}%
\bibitem [{\citenamefont {Patil}\ \emph {et~al.}(2014)\citenamefont {Patil},
  \citenamefont {Aycock}, \citenamefont {Chakram},\ and\ \citenamefont
  {Vengalattore}}]{Patil2014}%
  \BibitemOpen
  \bibfield  {author} {\bibinfo {author} {\bibfnamefont {Y.~S.}\ \bibnamefont
  {Patil}}, \bibinfo {author} {\bibfnamefont {L.~M.}\ \bibnamefont {Aycock}},
  \bibinfo {author} {\bibfnamefont {S.}~\bibnamefont {Chakram}}, \ and\
  \bibinfo {author} {\bibfnamefont {M.}~\bibnamefont {Vengalattore}},\
  }\href@noop {} {\bibfield  {journal} {\bibinfo  {journal} {arXiv:1404.5583}\
  } (\bibinfo {year} {2014})}\BibitemShut {NoStop}%
\end{thebibliography}
\end{document}